\documentclass[aps,pra,reprint,groupedaddress]{revtex4-1}

\usepackage{amsmath,amssymb,graphicx,hyperref}
\usepackage{array}
\usepackage[caption=false]{subfig}
\usepackage{bm}
\usepackage{amsfonts}
\usepackage{epsfig,epstopdf}
\usepackage{subfig}
\usepackage{url}

\def\Tr{{\rm Tr}}

\def\la{\left\langle}
\def\ra{\right\rangle}

\def\br{{\bf r}}

\def\bM{{\bf M}}

\def\bu{{\bf u}}

\def\hrho{\hat \rho}

\def\cH{{\cal H}}

\def\trace{{\rm Tr\ }}

\def\sinc{{\rm sinc}}

\def\be{\begin{equation}}
\def\ee{\end{equation}}
\def\ba{\begin{align}}
\def\nn{\nonumber\\}

\def\defeq{\buildrel \rm def \over =}

\def\b0{{\bf 0}}

\def\la{\left\langle}
\def\ra{\right\rangle}

\def\br{{\bf r}}

\def\bu{{\bf u}}

\def\bM{{\bf M}}

\def\ut{\underline{t}}

\def\bl{\bm{l}}

\def\brho{{\boldsymbol\rho}\,}
\def\cH{{\cal H}}

\def\Tr{{\rm Tr\ }}

\def\sinc{{\rm sinc}}

\def\be{\begin{equation}}
\def\ee{\end{equation}}
\def\ba{\begin{align}}
\def\nn{\nonumber\\}
\def\la{\langle}
\def\ra{\rangle}

\def\defeq{\buildrel \rm def \over =}

\def\Re{{\rm Re}}

\def\cR{{\cal R}}
\def\cB{{\cal B}}
\def\cD{{\cal D}}
\def\dl{d_\lambda}

\def\tdl{{\tilde d}_\lambda}
\def\tS{{\tilde S}}
\def\tF{{\tilde F}}
\def\tG{{\tilde G}}
\def\Dl{D_\lambda}
\def\sinc{{\rm sinc}\,}
\def\ud{\underline{d}}
\def\drho{\partial\hrho}
\def\pmu{\partial_\mu}

\begin{document}

\title{Quantum Limited Source Localization and Pair Superresolution in Two Dimensions under Finite Emission Bandwidth}

\author{Sudhakar Prasad}
\email{sprasad@unm.edu}
\affiliation{School of Physics
and Astronomy, University of Minnesota, Minneapolis, MN 55455}
\altaffiliation{Also at Department of Physics and Astronomy, University of New Mexico, Albuquerque, NM 87131}

\date{\today}

\pacs{(100.6640) Superresolution; (110.3055) Information theoretical analysis;
 (110.7348) Wavefront encoding; (110.1758)
Computational imaging; (270.5585) Quantum information and processing}

\begin{abstract}

Optically localizing a single quasi-monochromatic source to sub-diffractive precisions entails, in the photon-counting limit, a minimum photon cost that scales as the squared ratio of the width, $w$, of the optical system's point-spread function (PSF) and the sought localization precision, $d$, {\it i.e.,} as $\alpha(w/d)^2$. For sources with a finite emission-frequency spectrum, while the inverse quadratic scaling is expected to remain unchanged, the coefficient $\alpha$ must increase due to a degrading fidelity of localization as the imaging bandwidth increases and PSF undergoes a frequency-dependent widening. We specifically address how rapidly $\alpha$ must increase with increasing width of a flat-top spectral profile of emission of a point source being localized in two dimensions by an imager with a clear circular aperture by calculating quantum Fisher information (QFI), whose inverse yields the lowest possible unbiased-estimation variance of source-localization error. The novel use of prolate spheroidal wave functions as a basis for obtaining a solution of the eigenvalue problem of the single-photon density operator needed for the QFI calculation helps us develop the notion of an effective dimensionality of the continuous-state problem in terms of the associated space-bandwidth parameter. We subsequently extend our considerations of QFI to treat the finite-bandwidth pair superresolution problem in two dimensions, obtaining similar results. We also consider generalizations to emission power spectra of arbitrary profiles.
\end{abstract}

\vspace{-1cm}


\maketitle

\section{Introduction}\label{sec:intro}
Modern optical superresolution (OSR) imaging has drawn much interest over the past fifty years, starting with the pioneering modern work of Rushforth and Harris \cite{Rushforth68} on the role of noise in classical image restoration from spatially filtered images. Novel optical designs utilizing super-oscillating point-spread functions (PSFs) \cite{Berry06, Yuan16, Gbur19}, new metamaterial based super-lenses \cite{Durant06, Jacob06, Salandrino06, Liu07}, structured-illumination microscopy (SIM) \cite{Gustaffson00}, superresolution imaging flourescence imaging (SOFI) \cite{SOFI09}, superresolution imaging with quantum emitters employing quantum correlations in second \cite{Schwartz12} and higher orders \cite{Schwartz13, Monticone14, Israel17}, and SIM enhanced by quantum correlations of single emitters\cite{Classen17},  pushed at the theoretical limits of super-resolution in different ways. They all have practical limitations of one form or another, however, and achieve only moderate improvements by factors of order 2-3 even at very high signal-to-noise ratios. 

It was not until more recently that single-molecule localization imaging using uncorrelated photons from randomly photoactivated, well separated individual molecules \cite{Rust06} led to a qualitatively major advance in super-resolution, reaching ten to hundred fold improvement when compared to the classic Rayleigh-Abbe resolution limits. But these methods are limited to the biological domain where photoactivations and observations of only a subset of well separated fluorescent molecules are enabled, which only requires localization microscopy for each such sub-image, entailing a photon budget that follows an inverse quadratic dependence on the sought localization precision \cite{Thompson02,Ober04}. The final superresolved image only emerges when a large number of such source-localization-based subimages are carefully registered with respect to (w.r.t.) a fixed high-resolution grid and then superposed.  

The use of coherent detection techniques \cite{Roberts16,Yang16} have promised to enable qualitatively superior super-resolution of closely spaced point sources via quantum-correlated, optical centroid measuring states \cite{Tsang09, Schwartz13, Unternahrer18} and wavefront projections \cite{Tsang16, Nair16, Paur16, Rehacek17,Tham17,Chrostowski17,Zhou18,Tsang18,Tsang20}. These latter papers, led principally by the work of Tsang and collaborators \cite{Tsang16}, have provided the most fundamental, quantum mechanical estimation-theoretic limits of superresolution possible by {\it any} method and their realization for point-source imaging in domains including and, notably, beyond microscopy. In the photon counting limit, the variance for estimating the separation between a closely spaced, symmetrical pair of point sources using wavefront projections can, in principle, approach this quantum limit, with the corresponding photon cost scaling according to an inverse-square law w.r.t.~separation, rather than the inverse quartic law for intensity-based images \cite{Ram06,Prasad14}.   

Three recent papers by the present author \cite{YuPrasad18, PrasadYu19, Prasad20} have derived quantum estimation-theoretic limits on full three-dimensional (3D) localization and separation of a pair of incoherent point sources when the sources emit at a single wavelength. Coherent wavefront projections have been proposed and demonstrated as a way of realizing the lowest possible, quantum-mechanical bound on the variance of an unbiased estimation of the pair separation, as determined by the inverse of the quantum Fisher information (QFI) \cite{Helstrom76,Braunstein94,Paris09}. 

The projective wavefront coding approach to spatial point-source-pair OSR can be readily generalized to the time-frequency domain as well, as shown in Ref.~\cite{Donohue18} for a pair of Gaussian pulse forms with slightly different center frequencies for their spectra using Hermite-Gaussian time-frequency modes. But the calculation and possible realization of the quantum bounds on the spatial OSR problem when source emission has a finite optical bandwidth, a problem that combines experimentally relevant spatial and temporal characteristics in a single setting, has not been treated before.  The fundamental quantum bound on the variance of estimation of both the location and separation of the source pair by an imager is expected to degrade with increasing bandwidth of incoherent emission, since the imager's  PSF being optical-frequency dependent broadens. 

In this paper, we calculate the quantum estimation-theoretic fidelity for two problems of interest of finite-bandwidth emission in two dimensions - the transverse localization of a single point source w.r.t.~the optical axis and the separation of a pair of equally bright point sources that are symmetrically located w.r.t.~to the optical axis.  Assuming uniform incoherent emission over a finite bandwidth, with no emission outside it, we utilize the basis of one-dimensional (1D) prolate spheroidal wave functions (PSWFs) to calculate QFI for these two problems when the imaging pupil is a clear circular disk with perfect transmission, for which the PSF is of the Airy form \cite{Goodman96}.  Since, as previously noted \cite{Paur16,YuPrasad18}, in the photon counting limit the symmetrical pair OSR problem with a fixed midpoint of the pair separation vector and the single-source localization problem entail the same minimum estimation error, we expect to obtain similar results for the two problems. 

The use of PSWFs largely eliminates errors that would accrue from a direct numerical integration of the relevant integral eigenvalue equation based on a singular kernel function \cite{Bertero98}, while yielding important insights into the notion of an effective dimensionality \cite{Landau62,diFrancia69} of the continuous-state problem. The PSWF based approach, as we show in Ref.~\cite{Prasad20c}, also furnishes an excellent method for computing the quantum limits on superresolution imaging of spatially extended 1D and two dimensional (2D) sources.

\section{Quantum Bound on Source Localization with Single Photons}
Let a point source, which is located at position $\br$ in the plane of best focus, emit a photon into a uniformly mixed state of finite bandwidth $B\omega_0$ centered at frequency $\omega_0$ and let the photon be subsequently captured by an imaging system with aperture function $P(\bu)$. The state of such a photon may be described by the following single-photon density operator (SPDO):
\be
\label{rho}
\hrho = {1\over B}\int_\cB df\, |K_f\ra\la K_f|,
\ee
in which $f=(\omega-\omega_0)/\omega_0$ is the normalized frequency detuning, obtained by dividing the difference of the actual frequency, $\omega$, from the center frequency, $\omega_0$, by the latter. Correspondingly, the fractional detuning range, $\cB$, denotes the symmetrical interval, $-B/2<f<B/2$. Typical values of $B$ are expected to be small compared to 1. The wavefunction for the photon emitted into the pure state, $|K_f\ra$, of normalized frequency detuning $f$ and then captured by the imaging system has the following form in the system's exit pupil \cite{YuPrasad18}:   
\be
\label{wavefunction}
\la \bu| K_f\ra =  {1\over\sqrt{\pi}}P(\bu)\, \exp[-i2\pi (1+f)\bl\cdot\bu], 
\ee
where the pupil position vector $\bu$ is the true position vector normalized by dividing the latter by the characteristic spatial scale $R$ of the exit pupil.  For a circular aperture, we will take $R$ to be the radius of the exit pupil.  The symbol $\bl$ denotes the normalized transverse location vector of the point source, $\bl=\br/\delta$, obtained by dividing its physical position vector $\br$ by the characteristic Airy diffraction parameter, $\delta\defeq \lambda_0 z_I/R$, corresponding to the center optical wavelength, $\lambda_0=2\pi c/\omega_0$, and the distance $z_I$ of the image plane from the exit pupil. The parameter $\delta$ sets the Rayleigh resolution scale. In this section, we consider the minimum quantum limited variance of estimation of the distance, $l=|\bl|$, of the source from a known origin using a circular imaging pupil, assuming that the angle $\phi$ that $\bl$ makes with the $x$ axis may be estimated accurately in advance. 

The matter of how well we can localize a point source in the photon-counting limit can be treated by calculating the single-photon QFI w.r.t.~the source distance, $l$, from a fixed origin and then simply scaled up by multiplying the result with the observed number of photons. Such scaling is well justified for most thermal and other incoherent sources in nature because of their low mean photon emission number per coherence interval \cite{Goodman15}, $\delta_c <<1$, which thus may be regarded as emitting photons independently. The quantum state of $N$ independently emitted photons may be described by a tensor product of the density operators for the individual photons, so the overall QFI when $N$ independent photons are observed is simply $N$ times \cite{Tsang16,Liu20} that for a single photon. The same scaling holds for the classical Fisher information as well.

\subsection{General Expression for QFI}

The QFI per photon w.r.t.~a set of parameters, $\{\theta_1,\cdots,\theta_P\}$, on which SPDO has a differentiable dependence, is defined as the matrix \cite{Helstrom76} with elements that are the real part, denoted by the symbol, Re, of the following trace, denoted by the symbol, Tr:
\be
\label{QFIdef}
H_{\mu\nu} = \Re\, \Tr \left(\hrho\hat L_\mu\hat L_\nu\right),
\ee
where $\hat L_\mu$ is the symmetric logarithmic derivative of SPDO, $\hrho$, defined in terms of the partial derivative $\pmu\hrho$ w.r.t.~parameter $\theta_\mu$ by the relation,
\be
\label{SLD}
\partial_\mu \hrho = {1\over 2}\left(\hat L_\mu\hrho+\hrho\hat L_\mu\right).
\ee
By evaluating the trace in Eq.~(\ref{QFIdef}) in the basis of orthonormal eigenstates, $\{\lambda_i,\ |\lambda_i\ra\,|\,i=1,2,\ldots\}$ and calculating the partial trace over the null space of SPDO in terms of the partial trace over its range space, we may express $H_{\mu\nu}$ as \cite{YuPrasad18}, 
\begin{align}
\label{Hmn1}
H_{\mu\nu}=&4\sum_{i\in \cR}{1\over \lambda_i}\Re \langle \lambda_i|\partial_\mu \hat\rho\,\partial_\nu \hat\rho|\lambda_i\rangle]+2\sum_{i\in \cR}\sum_{j\in \cR}\Bigg[{1\over {(\lambda_i+\lambda_j)}}\nn
&-{1\over \lambda_i}-{1\over\lambda_j}\Bigg]\Re\langle \lambda_i|\partial_\mu \hat\rho|\lambda_j\rangle\langle \lambda_j|\partial_\nu \hat\rho|\lambda_i\rangle,
\end{align}
where $\cR$ denotes the space of values of the index of the eigenstates of SPDO associated with non-zero eigenvalues and the symbol $\partial_\mu$ denotes first-order partial derivative with respect to the parameter $\theta_\mu$. 

For the present problem of estimating a single parameter, $l$, we may drop the parameter labels as well as the operator $\Re$ everywhere.  By incorporating the $i=j$ terms from the double sum into the first sum in Eq.~(\ref{Hmn3}), we arrive at the following expression for QFI:
\ba
\label{Hmn3}
H=&\sum_{i\in \cR}{1\over \lambda_i}\left[4\langle \lambda_i|(\partial \hat\rho)^2|\lambda_i\rangle-3\la\lambda_i|\partial\hrho|\lambda_i\ra^2\right]\nn
+&2\sum_{i\neq j\in \cR}\left[{1\over (\lambda_i+\lambda_j)}-{1\over \lambda_i}-{1\over\lambda_j}\right]|\langle \lambda_i|\partial\hat\rho|\lambda_j\rangle|^2.
\end{align}
As we see from Eq.~(\ref{Hmn3}), evaluating QFI requires accurately computing the eigenstates and eigenvalues of SPDO given by Eq.~(\ref{rho}). 

\subsection{Eigenstates and Eigenvalues of SPDO }

In view of expression (\ref{wavefunction}), the overlap function of two single-photon states at two different frequency detunings $f,f'$ is given by the following pupil-plane integral over the normalized position vector, $\bu=\brho/R$:
\ba
\label{overlap}
O(f-f') &\defeq \la K_f|K_{f'}\ra\nn
          &= \int d^2 u |P(\bu)|^2\exp [i2\pi(f-f')\bl\cdot\bu].
\end{align}
For a circular clear pupil, for which $P(\bu)$ is simply $1/\sqrt{\pi}$ times the indicator function over the unit-radius pupil, the above integral may be evaluated in terms of Bessel function $J_1$ as \cite{Goodman96}
\be
\label{overlap1}
O(f-f') = {J_1(2\pi|f-f'|l)\over \pi|f-f'|l},
\ee
which reduces to 1 when $f\to f'$, as required by normalization of the single-photon states. The set of states, $\{|K_f\ra\}$, is clearly non-orthogonal.

Let $|\lambda\ra$ be an eigenstate of $\hrho$ of non-zero eigenvalue $\lambda$. Since $\hrho$ is supported over the subspace $\cH_B$ spanned by the basis $\{|K_f\ra,\ f\in \cB\}$, all its eigenstates with non-zero eigenvalues must also be fully contained in $\cH_B$. Consider therefore an expansion of $|\lambda\ra$ in this basis of form,
\be
\label{expansion}
|\lambda\ra = {1\over B} \int_\cB df' \, \dl(f') |K_{f'}\ra.
\ee
On substituting expressions (\ref{rho}) and (\ref{expansion}) for $\hrho$ and $|\lambda\ra$ into the eigenstate relation,
\be
\label{eigenrelation}
\hrho|\lambda \ra =\lambda |\lambda\ra,
\ee
and then equating the coefficients of each $|K_f\ra$ term on the two sides of the resulting equation, which is permitted due to the linear independence of these monochromatic single-photon states, we obtain the following integral equation for the coefficient function $\dl(f)$:
\be
\label{eigenrelation1}
{1\over B} \int_\cB O(f-f')\, \dl (f')\, df' = \lambda \dl(f).
\ee

 By defining the Fourier transform of $\dl(f)$ as
\be
 \label{FTcoeff}
\Dl (x)=\int_{-B/2}^{B/2} \dl(f)\exp(i2\pi f lx)\, df,\ \ x\in \cR,
\ee
we may transform Eq.~(\ref{eigenrelation1}) to the Fourier domain, as we show in Appendix A, re-expressing it as
\ba
\label{FTcoeff4}
\int_{-1}^1  \sqrt{1-x^{'2}}\,\sinc Bl(x-x')\,\Dl(x')\,&dx'={\pi \lambda\over 2} \Dl(x).\nn
& \ x\in \cR,
\end{align}
Note that without the square root inside the integrand Eq.~(\ref{FTcoeff4}) would be identical to the integral equation obeyed by the prolate spheroidal wave functions (PSWFs) first introduced by Slepian and Pollak \cite{Slepian61}. 

Let us expand $\Dl(x)$ in the complete orthogonal PSWF basis over the interval $(-1,1)$,
\be
\label{PSWFexpansion}
\Dl(x) = \sum_n d_n^{(\lambda)}\Psi_n(x;C),
\ee
where $C\defeq\pi Bl$ is the space-bandwidth parameter (SBP) of the associated PSWF problem. Substituting expansion (\ref{PSWFexpansion}) into Eq.~(\ref{FTcoeff4}), we can convert the original SPDO eigenvalue problem into a matrix eigenvalue problem of form,
\be
\label{FTcoeffM1}
\bM \ud^{(\lambda)}=\lambda \ud^{(\lambda)},
\ee
in which $\ud^{(\lambda)}$ denotes the column vector of coefficients,
\be
\label{columnvector}
\ud^{(\lambda)} = (d_0,d_1,\ldots)^T, 
\ee
with the superscript $T$ on a matrix denoting its simple transpose and the elements of the matrix $\bM$ are defined as the integral,
\be
\label{M}
M_{mn} = {2\over C}\int_{-1}^1 \sqrt{1-x^2}\, \Psi_m(x)\, \Psi_n(x)\, dx.
\ee
We relegate the details of this evaluation to Appendix A. 

The PSWFs alternate in parity,
\be
\label{PSWFparity}
\Psi_n(-x;C)=(-1)^n\Psi_n(x;C),
\ee 
and their associated eigenvalues $\lambda_n(C)$ are all positive and arranged in descending order, and obey the sum rule,
\be
\label{PSWFsum}
\sum_{n=0}^\infty \lambda_n^{(C)} = 2{C\over \pi},
\ee
with approximately $S\defeq\lceil 2C/\pi\rceil $ of these eigenvalues being close to $\min(2C/\pi,1)$ and the rest decaying rapidly toward 0 with increasing index value. Here the function $\lceil x\rceil$ takes the value 1 plus the integer part of $x$. The number $S$ is called the Shannon number, which was first introduced and discussed in the context of imaging by Toraldo di Francia \cite{diFrancia55, diFrancia69} as a measure of the effective number of degrees of freedom when a finite object is imaged with a finite-aperture imager. 

Since the PSWF $\Psi_n$ is either even or odd under inversion according to whether the index $n$ is even or odd, it follows that $M_{mn}$ is non-zero only if $m$ and $n$ are either both even or both odd. It then follows from Eq.~(\ref{FTcoeffM1}) that the set of coefficients $\{d_n^{(\lambda)}|n=0,1,\ldots\}$ separates into two subsets of coefficients, namely $\cD_e=\{d_n^{(\lambda)}|n=0,2,\ldots\}$ and $\cD_o=\{ d_n^{(\lambda)}| n=1,3,\ldots\}, $ that are only coupled within each subset. Correspondingly, in view of expansion (\ref{PSWFexpansion}) and parity-alternation property (\ref{PSWFparity}), the associated eigenfunctions $\Dl(x)$ are either even or odd under inversion, a fact that also follows directly from the form of the kernel of the integral equation (\ref{FTcoeff4}). For the two sets of even-order and odd-order coefficients, the matrix eigenvalue equation (\ref{FTcoeffM1}) may be solved quite efficiently for its eigenvalues and eigenvectors by truncating the size of the matrix at some finite but sufficiently high value $N$, {\it i.e.}, $0\leq m,n\leq N-1$. We evaluated integral (\ref{M}) by approximating the integral by a discretized Riemann sum and then using the Matlab routine {\it dpss} \cite{Percival98} that computes discrete sequences of the PSWFs for different values of SBP and sequence length on the interval $(-1,1)$. 

Due to the closeness of Eq.~(\ref{FTcoeff4}) to the integral equation obeyed by the PSWF, we expect there to be only a number of order $S$ of significantly large non-negative eigenvalues, $\lambda_p$, with the largest one being of order 1 and the successively smaller eigenvalues decreasing rapidly by many orders from one to the next. In other words, the nominal rank and the dimension of the range space of SPDO $\hrho$ are both expected to be of order $S$. This observation renders the problem numerically highly efficient, particularly when $C\sim 1$, for which the truncation index value, $N$, need not be greater than 10-20. These properties and the sum rule,
\be
\label{eig_sumrule}
\sum_{p=0}^\infty \lambda_p= 1,
\ee
obeyed by the eigenvalues of $\hrho$, since  ${\rm Tr}\,\hrho = 1$, were verified numerically. 

\subsection{Evaluation of QFI for 2D Source Localization}

By differentiaing expression (\ref{rho}) w.r.t.~$l$, we obtain
\be
\label{drho}
\partial\hrho={1\over B}\int df\, [\partial|K_f\ra\la K_f|+|K_f\ra\partial\la K_f|],
\ee
which, upon squaring and noting relation (\ref{overlap}), further yields 
\ba
\label{drho2}
(\partial\hrho)^2=&{1\over B^2}\int df\int df'\, [\partial|K_f\ra\la K_f|\partial|K_{f'}\ra\la K_{f'}|\nn
+\partial&|K_f\ra O(f-f')\, \partial\la K_{f'}|+|K_f\ra\partial\la K_f|\partial|K_{f'}\ra\la K_{f'}|\nn
+&|K_f\ra\partial\la K_f|K_{f'}\ra\partial\la K_{f'}|].
\end{align}
For notational brevity, we henceforth use the convention that $\partial$ only operates on the quantity immediately following it and have dropped explicit reference to the range, $(-B/2,B/2)$, of the frequency integrals. 

Next, taking the scalar product of the state vector $|K_{f'}\ra$ with expression (\ref{expansion}) for the eigenstate $|\lambda\ra$ and subsequently using the integral equation (\ref{eigenrelation1}) that the coefficients $d_\lambda (f)$ satisfies, we may show readily that 
\be
\label{Kf_lambda_matrixelement}
\la K_{f'}|\lambda_i\ra= \lambda_i d_i(f').
\ee
Use of expression (\ref{wavefunction}) for the wave function permits evaluation of the matrix element $\la K_{f'}|\partial|K_f\ra$ for a clear circular pupil for which $P(\bu)$ is simply $1/\sqrt{\pi}$ times its indicator function as
\ba
\label{KpK}
\la K_{f'}|\partial|K_f\ra=&-2i(1+f)\int_{u<1}\!\! \!\!d^2 u\cos\Phi_u\nn
                                      &\qquad\times \exp[-i2\pi (f-f')ul\cos\Phi_u]\nn
                                    =&-4\pi (1+f)\int_0^1 du\, u^2 J_1\big(2\pi (f-f')ul\big) \nn
                                    =&-{2(1+f)\over (f-f')l} J_2\big(2\pi (f-f')l\big)\nn
                                    =&(1+f)P(f-f'),\quad P(x)\defeq -2 {J_2(2\pi xl)\over xl}
\end{align}
in which $\Phi_u=\phi_u-\phi$ and we made successive use of the following identities for integrating first over the azimuthal angle, $\phi_u$, and then over the radial variable, $u$, of the pupil plane: 
\ba
\label{BesselIdentities}
\oint d\Phi\cos n\Phi \exp[\pm iz\cos(\Phi-\psi)] = &(\pm i)^n 2\pi \cos n\psi J_n(z);\nn
z^n J_{n-1} (z)=&{d\over dz}\left[z^n J_n(z)\right].
\end{align}
We can now evaluate the matrix element $\la\lambda_i|\partial\hrho|\lambda_j\ra$, with $\partial\hrho$  given by (\ref{drho}), by using relations (\ref{Kf_lambda_matrixelement}), (\ref{KpK}), and (\ref{expansion})  as,
\ba
\label{lambda_drho_lambda}
\la \lambda_i|\partial\hrho|\lambda_j\ra ={1\over B^2}&\int\int df \, df' (1+f)\big[\lambda_id_i(f')\, d_j(f) \nn
&+\lambda_j d_j(f')\, d_i(f)\big]\, P(f-f').
\end{align}

To evaluate the matrix elements $\la \lambda_i|(\partial\hrho)^2|\lambda_i\ra$, we first note from Eq.~(\ref{drho2}) that we need one more matrix element involving single-frequency emission states, namely $\partial \la K_f|\partial |K_{f'}\ra$, which we may evaluate as
\ba
\label{pKpK}
\partial \la K_f|\partial &|K_{f'}\ra =4\pi (1+f)(1+f')\nn
                                                    \times&\int_{u<1}d^2u\, u^2\cos^2\Phi_u \exp[-i2\pi(f-f')Bl\cos\Phi_u]\nn
= (2\pi)^2&(1+f)(1+f')\int_0^1du\, u^3\Big[J_0\big(2\pi (f-f')lu\big)\nn
                                      &\qquad\qquad+i^2J_2\big(2\pi (f-f')lu\big)\Big],
\end{align}
in which we used the identity, $2\cos^2\Phi_u=(1+\cos 2\Phi_u)$, and then used the first of the identities (\ref{BesselIdentities}) twice to reach the final equality. The indefinite integral of the first term in the integrand is known to be \cite{besint19}
\be
\label{BesselIdentity3}
\int dz\, z^3 J_0(z) = 2z^2 J_0(z) +z(z^2-4) J_1(z),
\ee
while the second term in the integrand may be evaluated immediately using the second of the identities (\ref{BesselIdentities}) for $n=3$. We obtain in this way the result,
\be
\label{pKpK2}
\partial \la K_f|\partial |K_{f'}\ra =(1+f)(1+f')\, Q(f-f'),
\ee
where the function $Q$ is defined by the relation
\ba
\label{Q}
Q(x)=&\Bigg[{2\over x^2l^2}\left(J_0\big(2\pi xl\big) -2{J_1\big(2\pi xl\big)\over 2\pi xl}\right)\nn
&+{2\pi\over xl}\left(J_1\big(2\pi xl\big)-J_3\big(2\pi xl\big)\right)\Bigg].
\end{align}  
In terms of the functions $O,\ P,$ and $Q$, we may express Eq.~(\ref{drho2}) as
\ba
\label{drho2a}
(\partial\hrho)^2=&{1\over B^2}\int df\int df'\, \big[\partial|K_f\ra (1+f')P(f'-f)\la K_{f'}|\nn
 &+\partial|K_f\ra O(f-f')\, \partial\la K_{f'}|\nn
 &+(1+f)(1+f')|K_f\ra Q(f-f')\la K_{f'}|\nn
 &+(1+f)|K_f\ra P(f-f')\partial\la K_{f'}|\big].
\end{align}
The matrix element $\la \lambda_i|(\partial\hrho)^2|\lambda_i\ra$ now follows from a repeated use of identity (\ref{Kf_lambda_matrixelement}) and expansion (\ref{expansion}), the latter yielding the relation,
\ba
\label{lambda_dKf}
\partial\la K_f|\lambda_i\ra^*=&\la \lambda_i|\partial|K_f\ra ={1\over B}\int df^{''}d_i(f^{''}) \la K_f^{''}|\partial|K_f\ra\nn
                                           =&{1\over B}(1+f)\int df^{''}  P(f-f^{''})d_i(f^{''}),
\end{align}
which can be evaluated efficiently by discretizing the integral as a Riemann sum that can be expressed as a matrix-vector product. 

\begin{figure}[htb]
\centerline{\includegraphics[width=0.9\columnwidth]{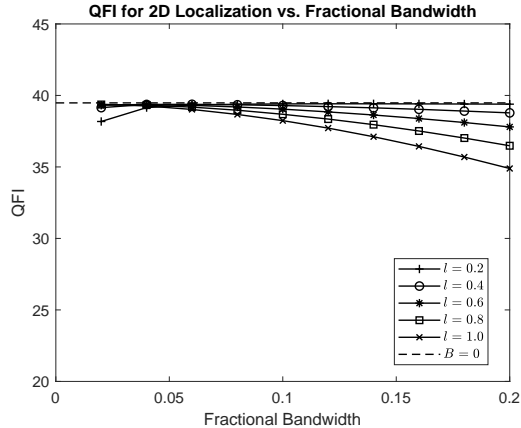}}
\vspace{-0.2cm}
\caption{Plot of QFI for estimating the distance, $l$, of a point source in the plane of Gaussian focus from the point of intersection of that plane with the optical axis vs. the fractional bandwidth, $B$, for different values of source distance $l$.}
\label{2DLocQFI_vs_B}
\end{figure}

In Fig.~\ref{2DLocQFI_vs_B}, we display numerically evaluated QFI for estimating the source location for a number of different values of its distance $l$ away from the {\it a priori} well determined axial point in the plane of Gaussian focus. The source distance, $l$, expressed in image-plane units of the Airy diffraction width parameter, $\lambda_0 z_I/R$, is allowed to vary in the sub-diffractive regime from 0.2 to 1.0. As expected and seen from the figure,  QFI decreases from the maximum theoretical zero-bandwidth value \cite{YuPrasad18} of $4 \pi^2$ as the fractional bandwidth, $B$, increases  but this decrease is rather gradual. Even for $l=1$, the maximum reduction of QFI at 20\% fractional bandwidth is no larger than about 10\%. The drop in QFI as $B$ varies between 0.02 and 0.04 for $l=0.2$ is presumably a numerical artifact, as we expect the localization QFI in this range to be quite close to the maximum value.  Since the minimum variance of unbiased estimation is the reciprocal of QFI \cite{Helstrom76}, the minimum quantum-limited error for estimating $l$ correspondingly increases with increasing $B$.

To see heuristically how bandwidth increase causes source-localization QFI to decrease, note that for monochromatic imaging ($B=0$) QFI is given by the expression \cite{YuPrasad18}, 
\be
\label{QFImono}
H=4\left[(\partial_l \la K_f|)\partial_l|K_f\ra-|\la K_f\partial_l|K_f\ra|^2\right],
\ee
which reduces, for an inversion symmetric aperture like the clear circular aperture, to four times the aperture average of the squared gradient w.r.t.~$l$ of the wavefront phase $\Psi$ in the aperture plane,
\be
\label{MeanSqPhaseGrad}
H  ={4\over \pi}\int d^2u P(\bu) (\partial_l \Psi)^2.
\ee
Since $\Psi=2\pi\,l\,u\,(1+f)\cos\Phi$ according to Eq.~(\ref{wavefunction}), Eq.~(\ref{MeanSqPhaseGrad}) evaluates for the clear circular aperture and monochromatic imaging ($f=0$) to the value $4\pi^2$. When the wavefunction $\la\bu|K_f\ra$, given by Eq.~(\ref{wavefunction}), is distributed over a finite bandwidth, the overall phase of any superposition of such wavefunctions gets scrambled, the more so, the larger the bandwidth, which increasingly reduces the mean-squared phase gradient over the aperture and thus $H$ with increasing bandwidth. This heuristic picture is supported by our quantitatively rigorous calculation of QFI based on the SPDO eigenvalues and eigenfunctions computed using PSWFs and displayed in Fig.~\ref{2DLocQFI_vs_B}.

\section{Quantum Bound on 2D Source-Pair Superresolution with Single Photons}
We will now evaluate QFI (\ref{Hmn3}) for estimating the separation of a symmetrical pair of closely spaced sources in a plane transverse to the optical axis of a circular-aperture imager. This calculation is closely related to the single-source localization QFI that we have considered so far. 

Consider a pair of equally bright incoherent point sources that are located at positions $\pm \bl$ with respect to the mid-point of their separation vector, which we choose to be fixed {\it a priori} at the origin. The SPDO for light emitted by the pair and transmitted through the imager to its pupil may be written as the integral
\be
\label{rho2}
\hrho={1\over 2B}\int_\cB\left[|K_{+f}\ra\la K_{+f}|+|K_{-f}\ra\la K_{-f}|\right]\, df                                                                               
\ee
in which, as for the localization problem, we take the detuning power spectrum of the imaging photon to be a top-hat function of fractional bandwidth $B$.   The state $|K_{\pm f}\ra$ is the pure monochromatic-photon state vector of fractional frequency $f$ emitted by the source located at $\pm \bl$, with its pupil-plane wave function of form,
\be
\label{wavefunction2}
\la \bu| K_{\pm f}\ra =  {1\over\sqrt{\pi}}P(\bu)\, \exp[\mp i2\pi (1+f)\bl\cdot\bu]. 
\ee
Because of the unit normalization of each of these states, $\la K_{\pm f}|K_{\pm f}\ra=1,$ expression (\ref{rho2}) has unit trace, $\trace(\hrho)=1$, as required. Also, the various pure-state overlap functions, for the case of a circular clear imaging aperture we are considering here, are real and equal in pairs,
\ba
\label{overlap2D}
\la K_{\pm f}|K_{\pm f'}\ra &= O(f-f');\nn
 \la K_{\pm f}|K_{\mp f'}\ra = &O(2+f+f'),
\end{align}
in terms of the function $O$ defined by relation (\ref{overlap1}). We now calculate the eigenvalues and eigenstates of SPDO (\ref{rho2}) in terms of which QFI (\ref{Hmn3}) is defined.

\subsection{Eigenvalues and Eigenstates of SPDO (\ref{rho2})}

Let an eigenstate of SPDO (\ref{rho2}) obeying the relation,
\be
\label{eigen2}
\hrho |\lambda\ra = \lambda|\lambda\ra,
\ee
have the expansion
\be
\label{expansion2}
|\lambda\ra = {1\over B}\int_\cB \left[d_+(f)|K_{+f}\ra+d_-(f)|K_{-f}\ra\right] df.
\ee
Substitution of expression (\ref{rho2}) and expansion (\ref{expansion2}) into eigenrelation (\ref{eigen2}) and use of relations (\ref{overlap2D}) for the various state overlap functions, followed by equating coefficients of the different monochromatic source states on the two sides of the resulting equation, yield the following pair of coupled integral equations for the coefficient functions, $d_\pm(f)$:
\ba
\label{IntEq2}
{1\over 2B}\int_\cB df'\, \big[&d_+(f')O(f-f')\nn
&+d_-(f') O(2+f+f')\big] = \lambda d_+(f);\nn   
{1\over 2B}\int_\cB df'\, \big[&d_+(f')O(2+f+f')\nn
&\qquad +d_-(f') O(f-f')\big] = \lambda d_-(f).
\end{align}

The two coupled equations in Eq.~(\ref{IntEq2}) may be decoupled by either adding them or subtracting one from the other as
\ba
\label{IntEq2Decoupled}
{1\over  2B}\int_\cB df'\, \left[O(f-f')+ O(2+f+f')\right]S_+(f') = &\lambda S_+(f);\nn   
{1\over 2B}\int_\cB df'\, \left[O(f-f')-O(2+f+f')\right]S_-(f') = &\lambda S_-(f),
\end{align}
where $S_+$ and $S_-$ are the sum and difference functions,
\be
\label{SA} 
S_+(f)=d_+(f)+d_-(f);\ \ S_-(f)=d_+(f)-d_-(f).
\ee

The two uncoupled equations (\ref{IntEq2Decoupled}) can be satisfied simultaneously by choosing either $S_+\neq 0,\ S_-=0$ or $S_+=0,\ S_-\neq 0$, corresponding, per Eq.~(\ref{SA}), to the choices $d_+(f)=\pm d_-(f)$. The nontrivial equation in each case may then be solved independently by using the same approach as for the 2D localization problem. Since the kernel functions, $[O(f-f')\pm O(2+f+f')]$, are not invariant under inversion, $f\to -f, \  f'\to -f'$, both even and odd PSWFs will be present in each such expansion, however. 

We first transform the problem to the Fourier domain,
\be
\label{FT2}
\tilde S_\pm(x) = \int_{-B/2}^{B/2} df \, \exp(i2\pi lxf)\, S_\pm(f),
\ee
and use the same $\delta$-function trick we used in going from Eq.~(\ref{AFTcoeff1}) to (\ref{AFTcoeff3}). Use of the Fourier shift theorem, which imples that the FT of the function $O(2+f)$ is simply $\exp(i4\pi lx)$ times the FT of the unshifted function, $O(f)$, we see that Eqs.~(\ref{IntEq2Decoupled}) transform to a pair of more convenient equations, which we can write more compactly as a single equation with its lower and upper signs corresponding to the two separate equations, 
\ba
 \label{IntEq2FTscaled}
 \int_{-1}^1& dx'\sqrt{1-x^{'2}}[\sinc Bl(x-x') \pm \exp(4\pi i lx')\nn
 &\times\sinc Bl(x+x')] \tS_\pm(x')=\pi\lambda\tS_\pm(x),\ \ x\in \cR.
 \end{align}

We may now substitute the spectral expansion (\ref{Aspectral_expansion}) of the sinc function and the expansion of the eigenfunctions $\tS_\pm (x)$ in terms of the PSWFs, namely
\be
\label{eigen2_expansion}
\tS_\pm(x)=\sum_{n=0}^\infty s^{(\pm)}_n \Psi_n(x;C),
\ee
into Eqs.~(\ref{IntEq2FTscaled}), then use the second of the orthogonality relations (\ref{APSWFnorm}), and finally equate the  coefficients of the individual PSWFs on both sides to convert those two integral equations into the following pair of matrix equations: 
\be
\label{eigen2_matrix_eq}
\sum_{n=0}^\infty \left[F_{mn}\pm (-1)^m G_{mn}\right] s^{(\pm)}_n=\lambda s^{(\pm)}_m,
\ee
in which the matrix elements $F_{mn}$ and $G_{mn}$ are defined as the integrals,
\ba
\label{FGmn}
F_{mn} = &{1\over C}\int_{-1}^1 \!\!dx'\sqrt{1-x^{'2}}\Psi_m(x';C)\Psi_n(x';C);\nn
G_{mn} = &{1\over C}\int_{-1}^1 \!\!dx'\sqrt{1-x^{'2}}\exp(4\pi ix')\,\Psi_m(x';C)\Psi_n(x';C).
\end{align}
To reach Eq.~(\ref{eigen2_matrix_eq}), we also used the parity-alternation property (\ref{PSWFparity}) of the PSWFs.

We now make use of the reality condition on the coefficient functions $d_\pm(f)$, or equivalently on their sum and difference functions, $S_\pm(f)$,  in the frequency domain. This condition requires that in the Fourier domain ($x$), the functions $\tS_\pm(x)$ obey the condition,
\be
\label{reality2}
\tS_\pm^*(x)=\tS_\pm(-x),
\ee
which upon substitution of expansion (\ref{eigen2_expansion}) and use of parity property (\ref{PSWFparity}) yields the equivalent condition,
\be
\label{reality2coeff}
s^{(\pm )*}_n=(-1)^n s^{(\pm)}_n.
\ee
In other words, the coefficients $s^{(\pm)}_n$ are alternately either purely real or purely imaginary, as the index $n$ ranges over all non-negative integer values. As such, we may express them in terms of real coefficients $t^{(\pm)}_n$ by the relation,
\be
\label{real_coeff}
s^{(\pm)}_n=i^n t^{(\pm)}_n.
\ee
A substitution of this relation into the eigenrelation (\ref{eigen2_matrix_eq}) yields the equivalent eigenrelation,
\be
\label{eigen2_matrix_eq_real}
\sum_{n=0}^\infty \left(\tF_{mn}\pm \tG_{mn}\right) t^{(\pm)}_n=\lambda t^{(\pm)}_m,
\ee
in which the matrix elements $\tF_{mn}$ and $\tG_{mn}$ are defined by the relation
\be
\label{real_matrix}
\tF_{mn}=i^{n-m} F_{mn},\ \tG_{mn}=i^{n+m}G_{mn}.
\ee
In view of the alternating parity of the PSWFs with changing order, the parity-even property of $\sqrt{1-x^{'2}}$ and of the integration range, the definitions (\ref{FGmn}) of the matrix elements, and since $\exp(4\pi i lx')$ is the sum of a real parity-even and an imaginary parity-odd part, we can see that ${\bf F}$ and ${\bf G}$ are symmetric matrices, $F_{mn}=0$ when the index difference $m-n$ is odd,  and $G_{mn}$ is purely real when $m+n$ is even and purely imaginary when $m+n$ is odd. It then follows that $\tF_{mn}$ and $\tG_{mn}$ defined by Eq.~(\ref{real_matrix}) are both real and symmetric. The eigenrelations (\ref{eigen2_matrix_eq_real}) are thus purely real equations involving symmetric system matrices, and are thus guaranteed to have real eigenvalues and orthogonal eigenvectors for non-degenerate eigenvalues.

We have numerically evaluated the eigenvalues and eigenvectors of the two matrices $(\tilde{\bf F}\pm\tilde{\bf G})$ by first calculating their matrix elements in terms of the discrete prolate spheroidal sequences discussed earlier, taking the latter to have a suffiiciently large length and truncating the matrices at some high but finite order of the PSWFs to ensure good accuracy. It helps, as with the localization problem, to know that only the largest $\mathcal{O}\lceil 2 C/\pi\rceil$ eigenvalues are sufficiently different from 0 to contribute significantly to QFI. In fact, for $C<<1$, which is the case of interest here, we ensure more than sufficient accuracy by truncating the matrix at order no larger than $15\times 15$ for which the smallest reproducibly computable eigenvalue has already dropped to a value more than fifteen orders of magnitude smaller than the largest one.

The orthogonality condition for the eigenvectors, $\la\lambda|\lambda'\ra=\delta_{\lambda \lambda'}$, can be shown, analogously to that for the localization problem, to be the same as Eq.~(\ref{Acolumn_orthogonality}), which for the column vector of real coefficients $t^{(\lambda)}_n$ is also the same,
 \be
\label{column_orthogonality2}
\ut^{(\lambda)\dagger}\ut^{(\lambda')}= {B\over l \lambda}\delta_{\lambda\lambda'},
\ee
where the superscript $(\lambda)$ labels the column vector corresponding to the eigenstate $|\lambda\ra$. Since the Hermitian transpose for a real column vector such as $\ut^{(\lambda)}$ amounts to its simple matrix transpose, we may renormalize each ordinary orthonormal eigenvector obtained from  a numerical evaluation of that eigenvector by an extra factor of $\sqrt{B/(l\lambda)}$.
 
\subsection{QFI Calculation}
We use expression (\ref{Hmn3}) for the evaluation of QFI for the single parameter of interest, the semi-separation parameter $l$. Unlike the localization problem, expression (\ref{rho2}) for SPDO is now more involved as it entails emission from two sources, rather than one. However, since we can work in the symmetric and anti-symmetric invariant subspaces of the DO, the two problems are rather analogous. In particular, we see that an eigenstate of SPDO in either of its $\pm$ range subspaces, which we denote as $\cH_B^{(\pm)}$, may be expressed as
\be
\label{eigen2_pm}
|\lambda^{(\pm)}\ra = {1\over B}\int d_+(f')\,\left(|K_{+f'}\ra\pm |K_{-f'}\ra\right) df',
\ee
with the notation $|\lambda^{(\pm)}\ra$ referring to an eigenstate belonging to the $\cH_B^{(\pm)}$ subspace. In view of this form, we can derive the relation,
\ba
\label{Kpf_lambda_pm}
\la K_{+f}|\lambda^{(\pm)}\ra =& {1\over B}\int d_+(f')\,\left[O(f-f')\pm O(2+f+f')\right] df'\nn
                                                 =&2\lambda^{(\pm)}d_+(f),
\end{align}
with the first relation following from a use of the overlap functions (\ref{overlap2D}) and the second from the eigenfunction relation (\ref{IntEq2}) in which we also used the fact that $d_-(f)=\pm d_+(f)$ in the two subspaces. We may similarly show that
\be
\label{Kmf_lambda_pm}
\la K_{-f}|\lambda^{(\pm)}\ra =2\lambda^{(\pm)} d_{-}(f)=\pm 2\lambda^{(\pm)}d_+(f).
\ee

The evaluation of the matrix elements $\la \lambda^{(\pm)}_i|\drho|\lambda^{(\pm)}_j\ra$ and $\la \lambda^{(\pm)}_i|(\drho)^2|\lambda^{(\pm)}_i\ra$ within each of the subspaces separately can now be carried out by differentiating expression (\ref{rho2}) with respect to $l$ first. The latter operation generates four terms, a pair of terms for each of the bilinear products, $|K_{+f}\ra\la K_{+f}|$ and $|K_{-f}\ra\la K_{-f}|$, inside the $f$ integral. Squaring $\drho$ then generates 16 terms inside a double frequency integral, for each of which terms one must evaluate the diagonal matrix element in an eigenstate $|\lambda^{(\pm)}_i\ra$. These calculations, although tedious, may be performed straightforwardly. Expressions (\ref{Kpf_lambda_pm}) and (\ref{Kmf_lambda_pm}) for the overlap functions greatly simplify these calculations, as we show in Appendix B, with the following results:
\ba
\label{MatElements2}
&\la\lambda_j^{(\pm)}|\drho|\lambda_i^{(\pm)}\ra\! =\! {2\over B^2}\!\iint\! df\, df' [P(f-f')\pm P(2+f+f')]\nn
&\times(1+f)\left[\lambda_i^{(\pm)}d_+^{(i)}(f)d_+^{(j)}(f')+\lambda_j^{(\pm)}d_+^{(j)}(f)d_+^{(i)}(f')\right];\nn
&\la\lambda_i^{(\pm)}|(\drho)^2|\lambda_i^{(\pm)}\ra={1\over 2B^2}\iint df\, df'\Big\{[O(f-f')\nn
&\qquad\pm O(2+f+f')]\la \lambda_i^{(\pm)}|\partial|K_{+f}\ra \la  \lambda_i^{(\pm)}|\partial|K_{+f'}\ra \nn
               &\qquad+4\lambda_i^{(\pm)}(1+f)[P(f-f')\pm P(2+f+f')]\nn
               &\qquad\times d_+^{(i)}(f)\la\lambda_i^{(\pm)}|\partial|K_{+f'}\ra \nn
               &\qquad+4\lambda_i^{(\pm)2}(1+f)(1+f')[Q(f-f')\nn
               &\qquad\pm Q(2+f+f')]d_+^{(i)}(f)d_+^{(i)}(f')\Big\},
\end{align}
where the functions $P$ and $Q$ have been defined earlier by Eqs.~(\ref{KpK}) and (\ref{Q}).  

The upper and lower signs in these expressions refer to the eigenstates drawn from the two subspaces, $\Omega_B^{(\pm)}$, respectively. What we also show in Appendix B is that any matrix element of form, $\la\lambda_j^{(\mp)}|\drho|\lambda_i^{(\pm)}\ra$, between any two states belonging to different subspaces vanishes identically,
\be
\label{MatElements3}
\la\lambda_j^{(\pm)}|\drho|\lambda_i^{(\mp)}\ra =0.
\ee
This allows both sums in expression (\ref{Hmn3}) to be evaluated separately over eigenstates belonging to the two different subspaces before adding their contributions to compute the total QFI.
\begin{figure}[htb]
\centerline{\includegraphics[width=0.9\columnwidth]{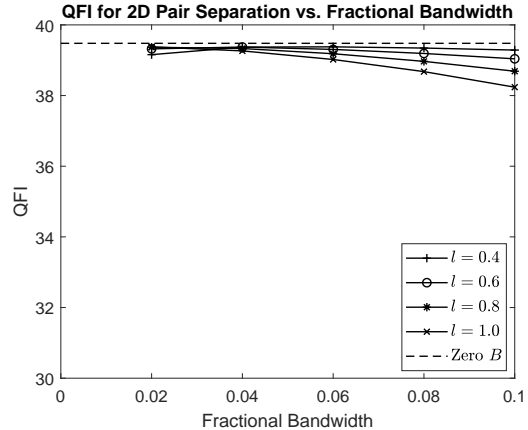}}
\vspace{-0.2cm}
\caption{Plot of QFI for estimating the semi-separation distance, $l$, of each point source from the pair centroid that has been fixed along the optical axis in the plane of Gaussian focus vs. fractional bandwidth, $B$, for different values of $l$.}
\label{2D_OSR_QFI_vs_B}
\end{figure}

\subsection{Numerical Results for QFI for 2D Pair OSR}
In Fig.~\ref{2D_OSR_QFI_vs_B}, we plot the value of QFI for estimating the separation of a symmetric source pair that is located in the transverse plane of Gaussian focus, with the origin in that plane fixed at the axial point that we take to be the pair's centroid. As the fractional bandwidth increases, QFI decreases much as it did for 2D localization of a single source that we treated in the previous section. However, even for 10\% fractional emission bandwidth and pair separations that are twice as large as the Airy parameter, QFI decreases to a value that is no more than 5\% below the maximum theoretical value of $4\pi^2$ for estimating the 2D pair separation distance for purely monochromatic emission. In other words, the maximum information that can be extracted about the pair separation remains rather robust with increasing emission bandwidth. 


\section{Realization of QFI via Low-order Zernike Projections}
We have noted previously \cite{YuPrasad18, PrasadYu19} that low-order Zernike wavefront projections furnish an entirely classical measurement protocol that can realize pair-superresolution QFI in the extreme limit of vanishing pair separation. The Zernike modes, being real and orthogonal over the unit disk, might also meet the optimality criteria laid out by Rehacek, {\it et al.} \cite{Rehacek17}, when extended to two dimensions with respect to a clear circular imaging aperture. We now show that the same protocol using the lowest four orders of Zernike polynomials, namely $Z_1, Z_2, Z_3,Z_4$ in Noll's notation \cite{Noll76}, works well even when the emission bandwidth of the sources is not particularly narrow and the source separation not too large. Since, due to the realness of the Zernike modes, the squared moduli of their normalized projections, which determine their probabilities, are the same for both the symmetric source-pair separation and single-source localization problems, identical results for Zernike-based classical FI (CFI) are obtained for both, provided the semi-separation distance in the former problem is identified, as we already have, with the source distance in the latter. 

The first four Zernikes are defined as the following functions of polar coordinates over the unit disk in the pupil plane:
\ba
\label{Z1234}
Z_1(\bu)=&{1\over \sqrt{\pi}};\ \ 
Z_2(\bu)={2\over\sqrt{\pi}}u\,\cos\phi_u;\nn 
Z_3(\bu)=&{2\over\sqrt{\pi}}u\,\sin\phi_u;\ \ 
Z_4(\bu)=\sqrt{3\over \pi}(1-2u^2).
\end{align}
The choice of the specific coefficients for these functions ensures that they have unit norm over the unit disk, {\it i.e.,} $\la Z_n|Z_n\ra=1$. The probability of observing an imaging photon in the $n$th Zernike mode, $P_n=\la Z_n|\hrho|Z_n\ra$, is the same whether $\hrho$ is given by Eq.~(\ref{rho}) or Eq.~(\ref{rho2}) for the two different problems considered in this paper. In view of form (\ref{wavefunction}) for the wavefunction, $\la \bu|K_f\ra$, we may express $P_n$ as
\be
\label{ProbZ}
P_n\!=\!{1\over \pi B}\int_{-B/2}^{B/2}\!\!\!\!\!\! df \left\vert\int\!\! P(\bu)\exp[-i2\pi(1+f)\bl\cdot\bu]\, Z_n(\bu) d^2u\right\vert^2.
\ee 
For the four Zernikes of interest here, we may calculate the corresponding probabilities as
\ba
\label{ProbZ1234}
P_1(l)=&{2\over B\pi l}\int_{x_-}^{x_+}dx{J_1^2(x)\over x^2};\nn
P_2(l)=&{8\over B\pi l}\cos^2\phi_l\int_{x_-}^{x_+}dx{J_2^2(x)\over x^2};\nn
P_3(l)=&{8\over B\pi l}\sin^2\phi_l\int_{x_-}^{x_+}dx{J_2^2(x)\over x^2};\nn
P_4(l)=&{96\over B\pi l}\int_{x_-}^{x_+}dx\Bigg[{J_0^2(x)\over x^4}+J_1^2(x)\Big({4\over x^6}-{1\over x^4}\nn &+{1\over 16 x^2}\Big)+J_0(x)J_1(x)\left({1\over2 x^3}-{4\over x^5}\right)\Bigg],
\end{align}
where $x_\pm$ are defined as
\be
\label{xpm}
x_\pm = 2\pi l\,(1\pm B/2).
\ee
We derived expressions (\ref{ProbZ1234}) by individually substituting the four Zernike polynomials (\ref{Z1234}) into Eq.~(\ref{ProbZ}), using the first of the Bessel identities in Eq.~(\ref{BesselIdentities}) to integrate over the angular coordinate $\phi_u$ in the unit disk, and then using the second of these identities and a third Bessel identity (\ref{BesselIdentity3}) to integrate over the radial coordinate $u$. The final step involved a simple scaling of the integration variable $f$ via the substitution $x=2\pi(1+f)l$. 

All of the integrals in Eq.~(\ref{ProbZ1234}) may in fact be evaluated in closed form. The values of the corresponding indefinite integrals, listed in the tables of Bessel integrals in Ref.~\cite{besint19} on pages 244 and 263, were used to express the requisite probabilities, $P_n(l), \ n=1,\ldots,4,$ in closed form. Their derivatives, $dP_n/dl$, on the other hand, are more simply calculated by noting that expressions (\ref{ProbZ1234}) depend on $l$ only through its presence in the denominator of the overall coefficient and in the integration limits, which renders this calculation quite simple when we use the identity,
\ba
\label{integral_identity}
{d\over dl}\left[{1\over l}\int_{b(l)}^{a(l)} f(x)\, dx\right]&=-{1\over l^2}\int_{b(l)}^{a(l)} f(x)\, dx\nn
&+{1\over l}\left[f(a) {da\over dl}-f(b){db\over dl}\right].
\end{align}

Based on the {\em observed} mode-projection probabilities and their derivatives, we can now calculate the classical FI for estimating the distance $l$. Since an imaging photon has the probability $\bar P=1-\sum_{n=1}^N P_n$ of being found in the {\em unobserved} modes, we can write down the full CFI \cite{VT68} per photon for estimating $l$ from projective measurements in the $N$ Zernike modes as
\be
\label{CFI}
F_N(l)=\sum_{n=1}^N {1\over P_n}\left({dP_n\over dl}\right)^2+{1\over \bar P}\left({d\bar P\over dl}\right)^2.
\ee

\begin{figure}[htb]
\centerline{\includegraphics[width=0.9\columnwidth]{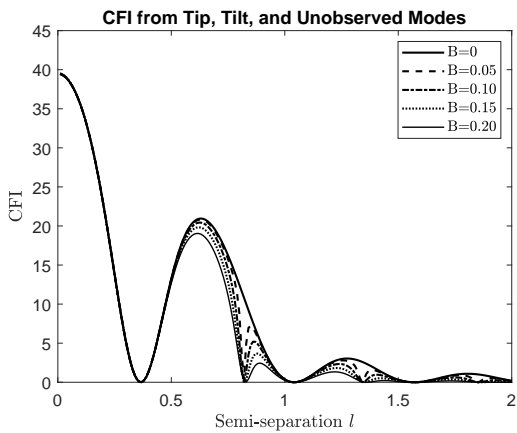}}
\vspace{-0.2cm}
\caption{Plot of CFI for estimating $l$ from wavefront projections into the tip-tilt modes, $Z_2$ and $Z_3$, vs. $l$ for a variety of values of the fractional bandwidth, $B=\Delta f/f_0$.}
\label{CFI_TipTilt}
\end{figure}

In Fig.~\ref{CFI_TipTilt}, we plot the numerically evaluated CFI for estimating $l$ when only projections into the tip and tilt modes, $Z_2,Z_3$, are observed and the remaining mode projections are not, for values of $l$ varying between 0 and 2 for five different values of the fractional bandwidth, $B$, namely 0, 0.05, 0.10, 0.15, and 0.20. As expected, the fidelity of estimation, represented by CFI, degrades with increasing bandwidth, since the diffraction induced image, whose width in the image domain is proportional to the wavelength, gets fuzzier with an increasing range of emission wavelengths. Note that the shorter the distance $l$, the less impact the bandwidth increase has on the value of tip-tilt CFI, which approaches the quantum FI in the limit of $l\to 0$, regardless of the value of $B$ even with observations in the tip and tilt modes alone. This behavior was noted earlier in Refs.~\cite{YuPrasad18,PrasadYu19} as arising from the fact that these tip and tilt modes are perfect matched filters for the $x$ and $y$ coordinates, respectively, of vector $\bl$ in this limit. The oscillatory behavior of the CFI curves with increasing $l$, with alternating local maxima and minima, on the other hand, have to do with the fact that at certain values of $l$, $dP_2/dl=dP_3/dl=0$ and consequently the first-order information provided by the tip and tilt modes alone about $l$ vanishes for those values.     

\begin{figure}[htb]
\centerline{\includegraphics[width=0.9\columnwidth]{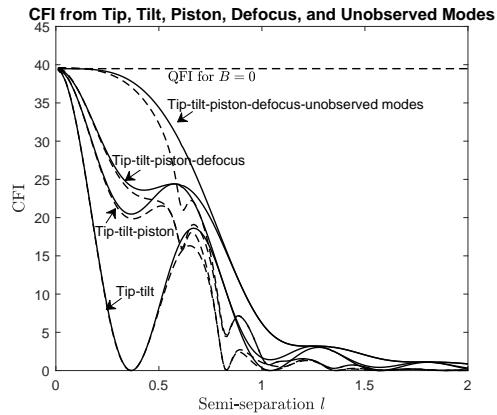}}
\vspace{-0.2cm}
\caption{Plot of CFI for estimating $l$ from wavefront projections into the tip-tilt, piston, and defocus modes, namely $Z_1,Z_2,Z_3,Z_4$, vs. $l$ for values 0 (solid lines) and 0.20 (dashed lines) of the fractional bandwidth, $B=\Delta f/f_0$. The bottom three pairs of closely-bunched curves capture the increase of CFI from partial contributions of the tip-tilt, piston, and defocus modes, while the top pair represent the total CFI from inclusion of the unobserved modes as well.}
\label{CFI_TipTiltPistonDefocus}
\end{figure}

The values of CFI increase with the inclusion of further Zernike modes, as Fig.~\ref{CFI_TipTiltPistonDefocus} demonstrates. In this figure, we plot the relative contributions of the various Zernike modes, starting with the tip and tilt modes for two different values of $B$, namely 0 and 0.2, which correspond to the same values of $B$ as for the outside pair of curves in Fig.~\ref{CFI_TipTilt}. The lowest pair of curves that are bunched together represent the tip-tilt contribution to CFI for the two values of $B$. The next higher closely paired curves display CFI for the same two values of $B$ when the contribution of the piston Zernike, $Z_1$, is added, while the second highest pair of curves exhibit CFI when the final Zernike mode, $Z_4$, often called the defocus mode, is also included. The very highest pair of curves represent the overall CFI when the contributions from these four Zernikes and all other unobserved modes are added together. In each curve pair, the higher, solid one corresponds to $B=0$ and the lower, dashed one to $B=0.20$. To avoid confusion, we have not displayed the dependence of CFI for the remaining three, intermediate values of $B$ also covered by Fig.~\ref{CFI_TipTilt}, but those dependences fall, as expected, between each pair of solid and dashed curves shown in Fig.~\ref{CFI_TipTiltPistonDefocus}. As we readily see, even adding the piston mode to tip-tilt mode projections greatly enhances CFI over a much larger range of separations than tip-tilt projections alone. 

\paragraph*{Discussion}
To gain some quantitative sense about the scale of resolution and number of photons that might be needed to achieve minimum acceptable estimation variances, let us consider the following example. A symmetrical pair of point sources are separated by $2l=0.4$, in Airy diffraction parameter units, emit at the center wavelength $\lambda_0=500$ nm, and produce geometrical images a distance $z_I=0.2$ m away from a thin spherical mirror of aperture radius, $R=0.1$ m.  In physical units, the pair separation has the value, $2l\,\delta=400$ nm. If the pair emission is observed in a 10\% fractional bandwidth ($B=0.1$), the values, per photon, of CFI calculated from observing projections into the tip-tilt Zernikes alone and tip-tilt-piston-defocus Zernikes alone are equal to 22.85 and 39.29, respectively, while QFI for $l=0.2$ and $B=0.1$ has the value 39.41, just a bit lower than the zero-bandwidth value of 4$\pi^2=39.48$ per photon. In other words, observing tip-tilt ($Z_2,Z_3$) projections alone can in principle achieve 58\% of the quantum upper bound on Fisher information (FI), while including piston and defocus mode ($Z_1,Z_4$) projections as well raises CFI to about 99.5\% of the quantum limit, making the latter classical limit essentially indistinguishable from the quantum upper bound. 

As for minimum standard deviations (SDs), $\sigma_l^{({\rm min})}$, for estimating $l$, assuming {\em unbiased} estimation, their quantum and classical lower limits are given by the square root of the reciprocals of QFI and CFI, respectively. For our specific example, we calculate these SDs on estimating $l$ to be 0.1593 and 0.1595 units per photon. For $N$ photons, since CFI and QFI both scale up by factor $N$ in the photon counting regime, the SDs are smaller by the factor $\sqrt{N}$. For $N=100$,  the minimum fractional error for estimating $l$ from the four lowest Zernike-mode projections is equal to $\sigma_l^{({\rm min})}/l=0.01593/0.2$, which is less than 8\%, making such estimations quite accurate even with just 100 photons. If finite detection errors, such as dark counts or additive noise are present, as is true for even the best photon counting detectors \cite{Hadfield09,Slussarenko19}, the minimum photon numbers needed for resolving such source pair would need to be higher.  

The mode projections, be they in the Zernike or another basis, can be acquired in the laboratory using a digital holographic mask \cite{Paur16} that encodes the modes in a Fourier optics set-up \cite{Goodman96, YuPrasad18}. A maximum-likelihood algorithm, as we also discussed in Ref.~\cite{YuPrasad18}, can then be used to recover the parameters of interest from the projection count data acquired using a photon-counting camera. 

\section{Concluding Remarks}

This paper has presented a theoretical analysis of the problems of quantum limited source localization and symmetrical point-source-pair separation in a single 2D object plane as the fractional bandwidth, $B$, of incoherent emission from the sources increases from zero and detection is limited only by photon counting noise. For both problems, the most important parameter that determines how the quantum estimation theroetic bound degrades with increasing fractional bandwidth, $B$, is the effective space-bandwidth parameter, $\pi B\ell$, where $\ell$, in units of the Airy diffraction parameter, is either the source distance from a fixed point when localizing a point source or the distance of either source from the {\it a priori} known midpoint of the line joining the pair of point sources when resolving the pair. In both cases, the fixed point was chosen without loss of generality to be the point at which the optical axis intersects the object plane taken to be the plane of Gaussian focus. 

The number of eigenstates of the imaging-photon density operator with eigenvalues significantly different from 0 and which thus significantly control the fundamental quantum limit on the minimum variance of any possible unbiased estimation of $l$ is of order $S\defeq \lceil 2Bl\rceil$, with that limiting minimum error variance increasing significantly only when $S$ greatly exceeds 1. We may regard $S$ as the effective dimensionality of the continuous-state eigenvalue problem for the single-photon density operator for a point source emitting incoherently in a finite bandwidth. We have used the machinery of prolate spheroidal wave functions to formulate and obtain a solution of the eigenvalue problem, with which we then calculated the quantum bound numerically for a clear circular imaging pupil, exhibiting the detailed manner in which the quantum error bound increases with increasing value of $S$ for the two problems. 

We have also shown that wavefront projections in the basis of Zernike modes can yield estimation fidelities approaching the quantum upper bound, even with a few low-order mode projections when the localization and pair separation distances are comparable to or smaller than the characteristic Rayleigh resolution scale. Including higher-order Zernike modes will surely reduce the gap between CFI and QFI for all values of such distances, but our recent work on quantum bounds for extended-source imaging has shown that this gap may not close fully even when {\em all} Zernike projections are included \cite{Prasad20c}. 

While this paper has considered in detail the simplest form of a uniformly distributed emission power spectrum over a finite bandwidth outside which it vanishes identically, any general integrable power spectrum may be treated by an adaptation of the present calculation, as we show without detailed numerical evaluations in Appendix C. For unimodal power spectra, such as Lorentzian and Gaussian power spectra, we can always identify an effective SBP of form $\pi Bl$, in which $B$ is of order full width at half maximum (FWHM) of the emission spectrum when expressed as a fraction of the center frequency of that spectrum. We expect the detailed calculations presented in this paper and conclusions drawn from them to hold qualitatively even for such general power spectra. 

Extensions of the finite-bandwidth QFI calculation to the axial dimension and pair brightness asymmetry for full 3D pair localization and separation will accord wavefront-projection-based superresolution techniques further value. Ultimately, however, these considerations will need to be generalized to finite sources with spatially non-uniform brightness distributions for a variety of low-light imaging applications. 

\acknowledgments
The author is grateful for the research facilities provided by the School of Physics and Astronomy at the U. of Minnesota where he has held the position of Visiting Professor for the last two years. This work was partially supported under a consulting agreement with the Boeing Company and by Hennepin Healthcare Research Institute under a research investigator appointment.
\appendix

\section{The Matrix Eigenvalue Problem}

Since, for a circular aperture, $O(f-f')$ is real and symmetric under inversion, as is the interval $\cB$, it follows from the eigenvalue equation (\ref{eigenrelation1}) that non-degenerate eigenvalues must be associated with coefficient functions that can be chosen to be real and are either even or odd under inversion, 
\be
\label{Aparity}
d_\lambda^*(f)=d_\lambda(f),\ \ d_\lambda(-f)=\pm d_\lambda (f).
\ee

The orthonormality of any two eigenstates, $|\lambda\ra$ and $|\lambda'\ra$,  corresponding to distinct eigenvalues, $\lambda\neq \lambda'$, namely $\la \lambda|\lambda'\ra=\delta_{\lambda\lambda'}$, may be expressed, in view of the expansion (\ref{expansion}) and the overlap function, $O(f-f')$, defined in (\ref{overlap}), as
\ba
\label{Aorthonormality}
\delta_{\lambda\lambda'}=\la \lambda|\lambda'\ra=&{1\over B^2}\iint_\cB df\, df' d_\lambda(f) d_{\lambda'}(f') O(f-f')\nn
                                    =&{\lambda'\over B}\int_\cB df\, d_\lambda(f) d_{\lambda'}(f),
\end{align}
in which we used the integral equation (\ref{eigenrelation1}) to arrive at the final expression.

Considering the coefficient function, $d_\lambda(f)$, to represent an amplitude spectrum with a sharp cut-off at $f=\pm B/2$, we can define the corresponding band-limited signal as its Fourier transform (FT) as 
\be
\label{AFTcoeff}
\Dl (x)=\int_{-B/2}^{B/2} \dl(f)\exp(i2\pi f lx)\, df,\ \ x\in \cR,
\ee
where $\cR$ denotes the set of all real numbers. The inverse FT (IFT) relation takes the form,
\be
\label{AIFTcoeff}
\dl(f)=l\Theta(B/2-|f|)\int_{-\infty}^\infty \Dl(x)\, \exp(-i2\pi flx)\, dx,
\ee
where $\Theta$ denotes the unit step function that takes the value 1 when its argument is positive and 0 when negative.
Taking the FT (\ref{AFTcoeff}) of  Eq.~(\ref{eigenrelation1}) after substituting the integral form (\ref{overlap}) for the overlap function into it and interchanging the order of the $\bu$ and $f$ integrals on the left-hand side (LHS) of the resulting equation, we may express it as
\be   
\label{AFTcoeff1}
{1\over \pi}\int_{0\le u\le 1} d^2u\, \Dl(\bu\cdot\bl/l)\, \sinc Bl(x-\bu\cdot\bl/l) = \lambda \Dl(x),
\ee
in which $\sinc x$ denotes the function $\sin(\pi x)/(\pi x)$. If we write Eq.~(\ref{AFTcoeff1}) as 
\ba
\label{AFTcoeff2}
{1\over \pi}&\int_{0\le u\le 1}\!\! \!\!d^2u\, \int_{-\infty}^\infty \!\!dx' \delta(x'-\bu\cdot\bl/l)\, \Dl(x')\, \sinc Bl(x-x')\nn
&=\lambda \Dl(x),
\end{align}
which follows from the projective property of the Dirac $\delta$ function, then interchange on the LHS the $x'$ and $\bu$ integrals and subsequently perform the angular part of the $\bu$ integral over the $(0,2\pi)$ period for the angle $\phi$ between $\bu$ and $\bl$ as
\ba
\label{Aangular_int}
\oint d\phi\,\delta(x'-u\cos\phi)= &{2\over u|\sin(\cos^{-1}(x'/u)|}\Theta(ul-|v'|)\nn
                                             = & {2\over \sqrt{u^2-x^{'2}}}\Theta(u-|x'|).
\end{align}
The remaining integral in expression (\ref{AFTcoeff2}) over the radial distance variable, $u$, over the range $(0,1)$ can now be carried out by substituting (\ref{Aangular_int}) into that expression, with the result
\be
\label{Aradial_int}
\int_0^1 du {u\over \sqrt{u^2-x^{'2}} }\Theta(u-|x'|) = \sqrt{1-x^{'2}}\Theta(1-|x'|),
\ee
and noting that only for $|x'| < 1$ does the unit step function on the LHS of integral (\ref{Aradial_int}) allow contribution to occur between the limits 0 and $1-x^{'2}$ on $x$. In view of the results (\ref{Aangular_int}) and (\ref{Aradial_int}), integral equation (\ref{AFTcoeff2}) simplifies to the form,
\ba
\label{AFTcoeff3}
{2\over \pi }\int_{-l}^l  \sqrt{1-x^{'2}}\sinc Bl(x-x')\,\Dl(x')\,&dx'=\lambda \Dl(x),\nn
& \ x\in \cR.
\end{align}

Any radially symmetric apodization of the pupil, {\it e.g.,} for certain astronomical applications \cite{Aime07}, will change the integrand of the radial integral (\ref{Aradial_int}) to include a squared apodization function $|P(u)|^2$. Such inclusion will change the value of the integral from that on the right-hand side (RHS) of Eq.~(\ref{Aradial_int}) and correspondingly the $(1-x^{'2})^{1/2}$ factor inside the integral equation (\ref{AFTcoeff3}). All subsequent considerations of the problem, apart from this change, would remain essentially the same, however, including the use of PSWFs to calculate the eigenvalues and eigenvectors.  

We may perform a spectral expansion of the sinc kernel in Eq.~(\ref{AFTcoeff3}) in terms of the corresponding PSWFs, denoted by $\Psi_n(x;C)$, $n=0,1,\ldots$, $C=\pi Bl$, as
\be
\label{Aspectral_expansion}
\sinc Bl(x-x') = {\pi\over C}\sum_{n=0}^\infty \Psi_n(x;C)\Psi_n(x';C),
\ee
with their norm defined according to the dual orthogonalization properties (\ref{APSWFnorm}). 

Since the PSWFs form a complete basis of orthonormal functions over the infinite line, we may expand the desired coefficient functions, $\Dl(x)$, uniquely in this basis as
\be
\label{APSWFexpansion}
\Dl(x) = \sum_n d_n^{(\lambda)}\Psi_n(x;C),
\ee
substitute this expansion into the integral equation (\ref{AFTcoeff3}), and use the linear independence of the PSWFs to express it as a matrix equation,
\be
\label{AFTcoeffM}
\sum_{m=0}^\infty M_{mn} d_m^{(\lambda)} = \lambda d_n^{(\lambda)}, \ n=0,1,\ldots,
\ee
in which the matrix element $M_{mn}$ is defined as the integral,
\be
\label{AM}
M_{mn} = {2\over C}\int_{-1}^1 \sqrt{1-x^2}\, \Psi_m(x)\, \Psi_n(x)\, dx.
\ee
To derive expression (\ref{AM}), we employed the following dual orthogonalization properties of PSWFs \cite{Slepian61,Simons11}:  
\ba
\label{APSWFnorm}
\int_{-\infty}^\infty dx\, \Psi_m(x;C)\Psi_n(x;C)=&\delta_{mn}; \nn
 \int_{-1}^1 dx\, \Psi_m(x;C)\Psi_n(x;C)=&\lambda_n^{(C)}\delta_{mn}.
\end{align}

Since the PSWF $\Psi_n$ is either even or odd under inversion according to whether the index $n$ is even or odd, it follows that $M_{mn}$ is non-zero only if $m$ and $n$ are either both even or both odd. It then follows from Eq.~(\ref{AFTcoeffM}) that the set of coefficients $\{d_n^{(\lambda)}|n=0,1,\ldots\}$ separates into two subsets of coefficients, namely $\cD_e=\{d_n^{(\lambda)}|n=0,2,\ldots\}$ and $\cD_o=\{ d_n^{(\lambda)}| n=1,3,\ldots\}, $ that are only coupled within each subset. Correspondingly, in view of expansion (\ref{PSWFexpansion}) and parity-alternation property (\ref{PSWFparity}), the associated eigenfunctions $\Dl(x)$ are either even or odd under inversion, a fact that also follows directly from the form of the kernel of the integral equation (\ref{AFTcoeff3}). 

\begin{figure}[htb]
\centerline{\includegraphics[width=0.9\columnwidth]{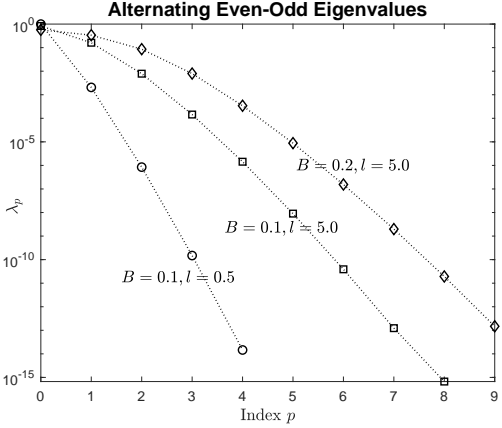}}
\vspace{-0.2cm}
\caption{Plots of the largest few eigenvalues for three different values for the pair $(B,l)$, namely $(0.1,0.5)$, $(0.1,5.0)$, and $(0.2,5.0)$, as indicated on the figure. The dotted lines connecting the successive discrete eigenvalues are provided merely to guide the eye.}
\label{2DLocEigenvalues}
\end{figure}

Figure \ref{2DLocEigenvalues} displays the largest few eigenvalues for three different pairs of values for $(B,l)$, as indicated. We immediately observe a rapid decrease of eigenvalues toward 0 with increasing index. These three pairs of values correspond to SBP, $C$, taking the values 0.157, 1.57, and 3.14. Note the several-orders-of-magnitude differences between successively smaller eigenvalues, those differences being more dramatic the smaller the value of $C$. The largest eigenvalue in each case is associated with an even eigenfunction.

\subsection{Eigenvector Orthonormality Condition}

Using Parseval's theorem for the FT, we may transform the orthonormality condition (\ref{Aorthonormality}) of any two eigenvectors of the single-photon DO to the Fourier domain as
\be
\label{AFTorthonormality1}
\int_{-\infty}^\infty \Dl(x)\, D_{\lambda'}^*(x)\, dx = {B\over l \lambda}\delta_{\lambda\lambda'},
\ee
When expansion (\ref{APSWFexpansion}) is substituted into the orthogonality relation (\ref{AFTorthonormality1}) and the first of the orthogonality relations (\ref{APSWFnorm}) for the PSWFs used, the following equivalent relation is obtained between the column vectors of coefficients, $d_n^{(\lambda)}$ and $d_n^{(\lambda')}$:
\be
\label{Acolumn_orthogonality}
\ud^{(\lambda)\dagger}\ud^{(\lambda')}= {B\over l \lambda}\delta_{\lambda\lambda'},
\ee
where the superscript $\dagger$ on a matrix denotes its Hermitian (or conjugate) transpose. 

The truncated version of $\bM$, as we have discussed earlier, will have $N$ different eigenvalues, $\lambda_p,\ p=0,\ldots, N-1$, and column eigenvectors, which we denote as $\ud_p,\ p=0,\ldots, N-1$. A typical numerical evaluation will generate the eigenvectors as being normalized to have unit Euclidean norm. Although their orthogonality is guaranteed by the symmetric, real form of $\bM$, meeting the orthonormality condition (\ref{Acolumn_orthogonality}) for the actual eigenvectors of the SPDO requires that such unit-Euclidean-norm eigenvectors be multiplied by $[B/(l\lambda_p)]^{1/2}$. Furthermore, since $d_\lambda(f)$ is a real function that is either even or odd under inversion, its complex FT, $\Dl(x)$, defined by relation (\ref{AFTcoeff}), is correspondingly either real (being two times the cosine FT of $\dl(f)$) or purely imaginary (being $2i$ times the sine FT of $\dl(f)$). If we sort our eigenvectors in the descending order of the associated eigenvalues, then we expect them to be alternately even and odd under inversion, much like the PSWFs themselves to which they are closely related.  Appending an extra overall factor of $i^p$ will ensure the real-imaginary alternation of successive eigenvectors, with successively increasing integer values of $p$. In view of these considerations and expansion (\ref{APSWFexpansion}), we may thus write
\be
\label{AReIm}
D_p(x)\defeq D_{\lambda_p}(x)=i^p \sqrt{B\over l\lambda_p}\ud_p^T\underline{\Psi}(x;C),\ \ p=0, 1, \ldots,
\ee
in which $\underline{\Psi}(x;C)$ denotes the column vector of the various PSWFs, namely
\be
\label{APsi}
\underline{\Psi}(x;C)=(\Psi_0(x;C)\ \Psi_1(x;C) \ \ldots)^T,
\ee
and $\ud_p$ are orthonormal real vectors of unit norm, $\ud_p^T\ud_{p'}=\delta_{pp'}$. 

Taking the IFT of  relation (\ref{AReIm}) w.r.t.~$x$, will yield the requisite coefficient function $d_\lambda(f)$ in the original frequency space labeled by $f$. This evaluation is greatly simplified when we make use of the remarkable property of the PSWFs that they are their own FT \cite{Percival98,Khare03,Wang17}, with a mere scaling of the argument, specifically that
\ba
\label{AFT_PSWF}
\int_{-\infty}^\infty &\exp(-i2\pi vf)\, \Psi_n(v/l;C)\, dv \nn
&=l\int_{-\infty}^\infty \exp(-i2\pi flx)\, \Psi_n(x;C)\, dx\nn
                                                          =&(-i)^n l  \sqrt{2\pi \over C\lambda_n^{(C)}}\Psi_n(2\pi fl/C; C)\,\Theta(1-2\pi |f|l/C)\nn
                                                          =&(-i)^n  \sqrt{2l\over B\lambda_n^{(C)}}\Psi_n(2 f/B; C)\,\Theta(B/2-|f|),
\end{align}
in which $\lambda_n^{(C)}$ are the eigenvalues associated with the PSWFs corresponding to SBP $C=\pi Bl$, was used to simplify certain arguments to arrive at the final expression. The unit step function factor ensures the band-limited property of the PSWFs.

In view of relation (\ref{AFT_PSWF}), the IFT of Eq.~(\ref{AReIm}) yields the coefficient function, $d_p(f)$, corresponding to the eigenvalue $\lambda_p$ as
\be
\label{AFcoeff} 
d_p(f) =i^p \sqrt{2\over \lambda_p}\ud_p^T\underline{\Psi'}(2f/B;C) \Theta(B/2-|f|),
\ee
where the column vector $\underline{\Psi}'$ is defined as 
\be
\label{APsi'}
\underline{\Psi}'=\left( {(-i)^0\Psi_0\over\sqrt{\lambda_0^{(C)}}}\  {(-i)^1\Psi_1\over\sqrt{\lambda_1^{(C)}}} \ \ldots\right)^T,
\ee
in which for brevity of notation we have omitted the argument list $2f/B;C$ of each PSWF. Since only either the even or odd order elements of the column vector $\ud_p$ are non-zero depending on whether $p$ is even or odd, respectively, the overall coefficient function $d_p(f)$, as given by Eq.~(\ref{AFcoeff}), is explicitly real for all values of $p$.

Note that both the eigenvalues, $\{\lambda_1,\lambda_2,\ldots\}$, and the eigenfunctions, $\{ d_1(f),d_2(f),\ldots\}$, over the normalized detuning interval, $-1< 2f/B< 1$, depend on the fractional bandwidth $B$ and pair separation $l$ only through their product, which is SBP, $C$. This fact can be quite useful in deriving any scaling laws w.r.t.~these two variables for QFI for which we will presently derive expressions.

\begin{figure}[htb]
\centerline{\includegraphics[width=0.9\columnwidth]{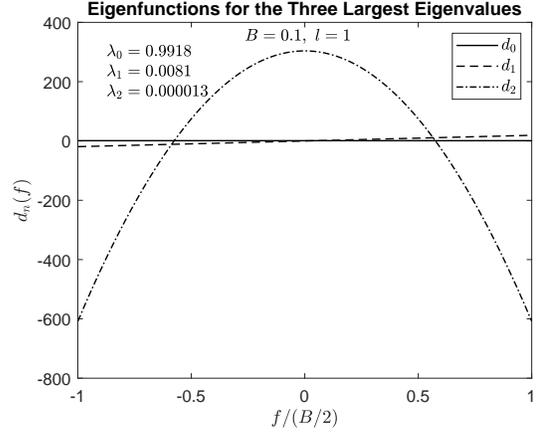}}
\vspace{-0.2cm}
\caption{Plots of the first three eigenfunctions, sorted according to decreasing eigenvalues, for the case of 10\% fractional bandwidth and source distance, $l=1$, from the axial point in the object plane. Their rapidly decreasing eigenvalues are listed on the figure.}
\label{2DLocEigenfunctions}
\end{figure}

The eigenfunctions associated with the three largest eigenvalues of the 2D localization SPDO (\ref{rho}) are plotted in Fig.~\ref{2DLocEigenfunctions} as functions of the normalized frequency, $\tilde f=f/(B/2)$, over the emission bandwidth for 10\% fractional bandwidth and source distance $l=1$ from the optical axis. For this case, SBP has the value $0.1\pi <1$, and is thus associated with only one significanly large eigenvalue, that associated with the even eigenfunction $d_0(f)$, when compared to all other eigenvalues. The eigenfunction $d_0(f)$  is essentially uniform with a value close to 1 throughout the frequency range, as shown by the solid curve on the plot that has a large vertical scale. The eigenfunctions associated with the second and third largest eigenvalues, which are, respectively, odd and even under inversion, have much larger values over most of the frequency range, even though their power, as measured by the largeness of their eigenvalues, is less than 1\% and 0.001\% of that for the highest-eigenvalue eigenfunction. 

\section{Matrix Elements of $\partial\hrho$ and $(\drho)^2$ for 2D OSR}

By differentiating $\hrho$ given by Eq.~(\ref{rho2}) w.r.t.~the pair separation $l$ once, we obtain
\ba
\label{drho_pair}
\drho={1\over 2B}\int_\cB  &\Big[\partial|K_{+f}\ra\la K_{+f}|+|K_{+f}\ra\partial\la K_{+f}|\nn
                                         &+\partial|K_{-f}\ra\la K_{-f}|+|K_{-f}\ra\partial\la K_{-f}|\Big]\, df                                                                               
\end{align}
Sandwiching this expression between an eigenket $|\lambda_i\ra$ and an eigenbra $\la \lambda_j|$,  we may express the resulting matrix element $\la\lambda_j|\drho|\lambda_i\ra$ as
\ba
\label{drho2_MatEl}
\la\lambda_j&|\drho|\lambda_i\ra={1\over B}\int_\cB  \Big[\la \lambda_j|\partial|K_{+f}\ra \lambda_id_+^{(i)}(f)\nn
&+\lambda_j d_{+}^{(j)}(f)\partial\la K_{+f}|\lambda_i\ra+\pi_i\lambda_id_+^{(i)}(f)\la \lambda_j|\partial| K_{-f}\ra \nn
&+\pi_j\lambda_jd_+^{(j)}(f)\partial\la K_{-f}|\lambda_i\ra\Big]\, df,                                                                             
\end{align}
in which $\pi_i$, taking the values $\pm 1$, denotes the parity of the eigenstate $|\lambda_i\ra$. 
To arrive at Eq.~(\ref{drho2_MatEl}), we used an analog of identity (\ref{Kf_lambda_matrixelement}) obtained by taking the inner product of eigenket expansion (\ref{expansion2}) for the $i$th eigenstate with $\la K_\pm(f')|$ and then using the integral equations (\ref{IntEq2}) to simplify the final result, namely
\be
\label{Kf_lambda2_MatEl}
\la K_\pm(f')|\lambda_i\ra=2\lambda_i d_\pm^{(i)} (f)= 2\pi_i\lambda_i d_+^{(i)}(f).
\ee
Another use of expansion (\ref{expansion2}), along with the definition (\ref{KpK}) of $P(f-f')$, simplifies the remaining matrix elements in Eq.~(\ref{drho2_MatEl}),
\ba
\label{drho2_MatEl2}
&\la\lambda_j|\drho|\lambda_i\ra={1\over B^2}\iint df\,df' (1+f) \Big\{\lambda_id_+^{(i)}(f)\, d_+^{(j)}(f')\nn
&\times\big[P(f-f') +\pi_j P(2+f+f')\big]\nn
&+\lambda_jd_+^{(j)}(f)\, d_+^{(i)}(f')\big[ P(f-f') +\pi_i P(2+f+f')\big]\nn
&+\pi_i\pi_j\lambda_id_+^{(i)}(f)\, d_+^{(j)}(f')\big[ P(f-f') +\pi_j P(2+f+f')\big]\nn
&+\lambda_jd_+^{(j)}(f)\, d_+^{(i)}(f')\big[ P(f-f') +\pi_i P(2+f+f')\big]\Big\},
\end{align}
where we have dropped the limits on the frequency integrals for bervity of notation. When the states $|\lambda_i\ra$ and $|\lambda_j\ra$ have opposite parities, {\it i.e.}, $\pi_i\pi_j=-1$, then the first pair of terms cancel the second pair of terms in Eq.~(\ref{drho2_MatEl}) exactly, while for the case of identical-parity states, $\pi_i=\pi_j$, the two pairs of terms add identically,
\ba
\label{dhro2_MatEl3}
\la\lambda_j^\pm|\drho|\lambda_i^\mp\ra=&0;\nn
\la\lambda_j^\pm|\drho|\lambda_i^\pm\ra=&{2\over B^2}\iint df\,df' (1+f) \nn
\times[P(f-f')& +\pi_i P(2+f+f')]\nn
\times\Big[\lambda_id_+^{(i)}(f)\, &d_+^{(j)}(f')+\lambda_jd_+^{(j)}(f)\, d_+^{(i)}(f')\Big].
\end{align}

By squaring Eq.~(\ref{drho_pair}) and then using the matrix elements defined via Eqs.~(\ref{overlap2D}), (\ref{KpK}), and (\ref{pKpK2}), we may express $(\drho)^2$ as
\ba
\label{drhosq2}
(\drho)^2=&{1\over 4B^2}\iint df\, df'\Big\{ (1+f) P(f-f')\nn
&\times\big[|K_{+f}\ra\partial\la K_{+f'}|+ |K_{-f}\ra\partial\la K_{-f' }|+{\rm h.a.}\big]\nn
&+(1+f)P(2+f+f')\nn
&\times\big[|K_{+f}\ra\partial\la K_{-f'}|+|K_{-f}\ra\partial\la K_{+f'}|+{\rm h.a.}\big]\nn
+ O&(f-f')\big[\partial|K_{+f}\ra\partial\la K_{+f'}|+\partial|K_{-f}\ra\partial\la K_{-f'}|\big]\nn
+O&(2+f+f')\big[\partial|K_{+f}\ra\partial\la K_{-f'}|+\partial|K_{-f}\ra\partial\la K_{+f'}|\big]\nn
+(&1+f)(1+f')Q(f-f')\nn
&\times\big[|K_{+f}\ra\la K_{+f'}|+|K_{-f}\ra\la K_{-f'}|\big]\nn
+(&1+f)(1+f')Q(2+f+f')\nn
&\times\big[|K_{+f}\ra\la K_{-f'}|+|K_{-f}\ra\la K_{+f'}|\big],
\end{align}
where the symbol h.a.~denotes the Hermitian adjoint of the terms preceding the symbol inside the square brackets. By taking the diagonal matrix element of expression (\ref{drhosq2}) in the state $|\lambda_i\ra$, and using identities (\ref{Kpf_lambda_pm}) and (\ref{Kmf_lambda_pm}) wherever possible, along with the relation,
\ba
\label{pm_identity}
\la \lambda_i|\partial|K_{+f}\ra=&{(1+f)\over B}\int d_+^{(i)}(f')\big[P(f-f')\nn
                                                  &\qquad+\pi_i P(2+f+f')\big]\nn
                                                =&\pi_i\la \lambda_i|\partial|K_{-f}\ra,
\end{align}
that follows from expansion (\ref{eigen2_pm}) and definition (\ref{KpK}) of the function $P$, we may, after a minor rearrangement of terms, express that element in the form,
\ba
\label{drhosq2_MatEl}
\la\lambda_i|(\drho)^2&|\lambda_i\ra={1\over 2B^2}\iint df\, df'\Big\{\big[O(f-f')\nn
&\quad+\pi_iO(2+f+f')\big]\la \lambda_i|\partial|K_{+f}\ra\la\lambda_i|\partial| K_{+f'}\ra\nn
&+4 \lambda_i(1+f) \big[P(f-f')+\pi_iP(2+f+f')\big]\nn
&\quad\times d_+^{(i)}(f)\la \lambda_i|\partial|K_{+f'}\ra\nn
&+4\lambda_i^2(1+f)(1+f')\big[Q(f-f')\nn
&\quad+\pi_i Q(2+f+f')\big]d_+^{(i)}(f)d_+^{(i)}(f')\Big\}.
\end{align}

\section{2D Localization under a More General Emission Spectrum}
For a single point source at distance $l$ w.r.t.~a fixed origin on the optical axis of the imager and emitting photons in a finite-bandwidth detuning spectrum of normalized power density function $W(f)$, the single-photon density operator takes the form,
\be
\label{Grho}
\hrho = \int_{-\infty}^\infty |K_f\ra\la K_f|\, W(f)\, df,\ \ \ \int^\infty_{-\infty} W(f)\, df=1.
\ee

We may now take an eigenstate $|\lambda\ra$ of non-zero eigenvalue $\lambda$ to be of form,
\be
\label{Gexpansion}
|\lambda\ra = \int \dl(f') \,  |K_{f'}\ra W(f')\, df', 
\ee
for which, following an identical set of steps as that which yields the integral equation (\ref{eigenrelation1}), we may derive the following integral equation for the coefficient function $\dl(f')$:
\be
\label{Geigenrelation1}
\int O(f-f')\, W(f')\,\dl (f')\, df' = \lambda \dl(f).
\ee

Taking the FT, defined as
\be
\label{GFTcoeff}
\tdl(v)=\int_{-\infty}^\infty \dl(f)\,\exp(i2\pi fv)\, df,
\ee
of  Eq.~(\ref{Geigenrelation1}) after substituting the integral form (\ref{overlap}) for the overlap function into it and interchanging the order of the $\bu$ and $f$ integrals on the LHS of the resulting equation, we may express it as
\be   
\label{GFTcoeff1}
{1\over \pi}\int_{-\infty}^\infty df' W(f')\,\dl(f')\, \int_{0\le u\le 1} \!\!\!\!\!\!\!d^2u\, \delta(v-\bu\cdot\bl)  = \lambda \tdl(v),
\ee
in which the Dirac $\delta$ function results from use of the identity,
\be
\label{DiracDelta}
\int_{-\infty}^\infty df\, \exp[i2\pi f(v-\bu\cdot\dl)]=\delta(v-\bu\cdot\bl).
\ee
That $\delta$ function is easily integrated, as we saw earlier in Section 2, over the unit disk in the $\bu$ plane to yield
\be
\label{DiracDeltaInt}
\int_{0\le u\le 1} d^2u\, \delta(v-\bu\cdot\bl)={2\over l^2}\sqrt{l^2-v^2}\Theta(l-|v|).
\ee
When result (\ref{DiracDeltaInt}) is substituted into the integral equation (\ref{GFTcoeff1}) for the eigenfunction and the convolution theorem of the FT is applied, the latter reduces to the form,
\be
\label{GFTcoeff2}
{2\sqrt{l^2-v^2}\over \pi l^2}\Theta(l-|v|)\int_{-\infty}^\infty \tdl(v')\tilde W(v-v')\, dv' = \tdl(v).
\ee
Since the LHS of Eq.~(\ref{GFTcoeff2}) vanishes for $|v|>l$, so must its RHS, namely $\tdl(v)$, and so the integral of the LHS can be restricted to the interval $(-l,l)$,
\be
\label{GFTcoeff3}
{2\sqrt{l^2-v^2}\over \pi l^2}\int_{-l}^l \tdl(v')\tilde W(v-v')\, dv' = \tdl(v),\ \ |v|<l.
\ee

\paragraph*{Symmetric Detuning Power Spectrum}

Equation (\ref{GFTcoeff3}) immediately implies that whenever the detuning spectrum $W(f)$ - and thus its FT $\tilde W(v)$ - is even under inversion, purely even and odd parity eigenfunctions are possible. Assuming such an even-parity spectrum, we may express the even and odd parity families of eigenfunction solutions of the integral equation (\ref{GFTcoeff3}) in terms of cosine and sine Fourier series, respectively, as
\ba
\label{Geigenfunctions}
\tdl^{(+)}(v)=&\sum_{m=0}^\infty d^{(+)}_m {1\over \sqrt{g_m l}}\cos(\pi mv/l)\, (1-v^2/l^2)^{1/4};\nn   
\tdl^{(-)}(v)=&\sum_{m=1}^\infty d^{(-)}_m {1\over \sqrt{l}}\sin(\pi mv/l)\, (1-v^2/l^2)^{1/4};
\end{align}
where $g_m=2$ for $m=0$ but 1 for $m\geq 1$. With the normalization factors inside the square root in these sums, the Fourier basis functions are orthonormal over the interval $(-l,l)$ of support of the coefficient functions $\tdl^{(\pm)}(v)$. 

Inserting the overall factors of $(1-v^2/l^2)^{1/4}$ in the solutions (\ref{Geigenfunctions}) implies a real, symmetric kernel for the integral equation  (\ref{GFTcoeff3}). Correspondingly, when expressions (\ref{Geigenfunctions}) are substituted into Eq.~(\ref{GFTcoeff3}) and the orthonormality of the basis functions is used, we obtain two families of linear equations for the coefficients $d^{(\pm)}_m$, each of which may be conveniently expressed as the product of a symmetric system matrix $\bM^{(\pm)}$ with the column vector of those coefficients, $\ud^{(\pm)}=(d^{(\pm)}_1,d^{(\pm)}_2,\ldots)^T$,
\be
\label{GMatEq}
\bM^{(\pm)} \ud^{(\pm)}=\lambda \ud^{(\pm)}.
\ee
The two system matrices $\bM^{(\pm)}$ have the elements,
\ba
\label{GMatEl}
M^{(+)}_{mn} = & {2\over \pi\sqrt{g_m g_n}}\iint_{-1}^1dx\, dx' (1-x^2)^{1/4}(1-x^{'2})^{1/4}\nn
&\quad\times\cos(\pi mx)\cos(\pi nx') \tilde W\big(l(x-x')\big);\nn
M^{(-)}_{mn} = & {2\over \pi}\iint_{-1}^1dx\, dx' (1-x^2)^{1/4}(1-x^{'2})^{1/4}\nn
&\quad\times \sin(\pi mx)\sin(\pi nx') \tilde W\big(l(x-x')\big),
\end{align}
where the integrals have been expressed in terms of the scaled variables, $x=v/l, \ x'=v'/l$.

The normalization condition, $\la\lambda|\lambda'\ra=\delta_{\lambda\lambda'}$, for the eigenstates within each of the two fixed-parity families, becomes an integral in the frequency domain, 
\be
\label{Gorthonormality}
\int_{-\infty}^\infty df\, d_\lambda(f)\, d_{\lambda'}(f) \, W(f)={1\over \lambda}\delta_{\lambda\lambda'},
\ee
as can be shown quite analogously to how condition (\ref{Aorthonormality}) was derived for the uniform emission problem earlier in Appendix A. Equivalently, by noting that the coefficient functions $\dl(f),\dl^{'}(f')$ are real and substituting the inverse Fourier-transform relation,
\be
\label{GIFT}
\dl(f)=\int_{-l}^l \tdl(v)\, \exp(-i2\pi fv)\, dv,
\ee
and its complex conjugate evaluated for eigenvalue $\lambda'$ into condition (\ref{Gorthonormality}), we may, with the help of integral equation (\ref{GFTcoeff3}), express that condition in the Fourier domain as
\be
\label{GFTorthonormality}
\int_{-1}^1 \tdl(lx)\, \tilde{d}_{\lambda'}^*(lx) \, (1-x^2)^{-1/2} dx= {2\over\pi \lambda^2l^2}\delta_{\lambda\lambda'}.
\ee
When we substitute expansions (\ref{Geigenfunctions}) into Eq.~(\ref{GFTorthonormality}) and use the orthogonality relations,
\ba
\label{FbasisOrtho}
\int_{-1}^1\cos (\pi mx)\,\cos(\pi n x)\, dx=&g_m\delta_{mn},\ \ m,n=0,1,2,\ldots;\nn
\int_{-1}^1\sin (\pi mx)\,\sin(\pi n x)\, dx=&\delta_{mn},\ \ m,n=1,2,\ldots,
\end{align} 
we obtain the following equivalent condition on the vectors of associated coefficients:
\be
\label{GFTorthoMat}
\ud^{(\pm)\dagger} \ud^{\,'(\pm)}={2\over \pi\lambda^2 l}\delta_{\lambda\lambda'}.
\ee
Note that if we assume the coefficient functions $\dl^{(\pm)}(f)$ to be real in frequency space, then the coefficients of the associated cosine and sine Fourier series (\ref{Geigenfunctions}) defining their Fourier transforms, $\tdl^{(\pm)}(v)$, must be real and purely imaginary, respectively,
\be
\label{ReImCoeff}
d_m^{(+)*}=d_m^{(+)},\ \ d^{(-)*}_m=-d_m^{(-)}.
\ee

With the eigenrelation in matrix form (\ref{GMatEq}), we can now calculate the eigenvalues and eigenfunctions using any standard numerical linear-algebra routine and subsequently employ the approach discussed in Sec.~IV to calculate QFI for a general symmetric detuning power spectrum $W(f)$. An analogous generalization applies to the symmetric pair-OSR problem as well.

\end{document}